\documentstyle[12pt]{article}
\begin{document}

\centerline{\Large\bf Images of very high energy cosmic ray sources} 
\vskip 0.6truecm
\centerline{\Large\bf in the Galaxy} 
\vskip 0.6truecm
\centerline{\large\bf I. A source towards the Galactic Centre}

\vskip 1.5truecm
\centerline{\large W. Bednarek, M. Giller and M. Zieli\'nska}

\vskip 0.7truecm
\centerline{Division of Experimental Physics, University of \L \'od\'z,}
\centerline{ul. Pomorska 149/153, 90-236 \L \'od\'z, Poland}

\vskip 1.truecm

\newpage

\centerline{Abstract}

\vskip 0.7truecm
Recent analyses of the anisotropy of cosmic rays at
$10^{18}$~eV (the AGASA and SUGAR data) show significant excesses from  
regions close to the Galactic Centre and Cygnus. Our aim is to check whether 
such anisotropies can be caused by single sources of charged particles. We 
investigate propagation of protons in two models of the Galactic regular magnetic 
field (with the irregular component included) assuming that the particles are 
injected by a short lived discrete source lying in the direction of the Galactic 
Centre. We show that apart from a prompt image of the source, 
the regular magnetic field may cause delayed images
at quite large angular distances from the actual source direction. The image is 
strongly dependent on the time elapsed after ejection of particles and it is
also very sensitive to their energy.
For the most favourable conditions for particle acceleration by a young pulsar
the predicted fluxes are two to four order of magnitudes higher than that observed.
The particular numbers depend strongly on the Galactic magnetic field model adopted
but it looks that a single pulsar in the Galactic Centre could be responsible for the 
observed excess.

\newpage

\section{Introduction}

Recent analysis of the AGASA data shows anisotropy in arrival directions of 
cosmic rays with energies $10^{17.9}-10^{18.3}$~eV, with excesses from two
directions near the Galactic Centre (4.5$\sigma$) and the Cygnus region
(3.9$\sigma$) (Hayashida et al.~1999). The existence of a point like
excess at $\sim 7.5^{\rm o}$ from the Galactic Centre (GC) has been confirmed
by the analysis of the SUGAR data (Bellido et al.~2001). 
Hayashida et al. suggested that
such point like excesses might be caused by relativistic neutrons which are
able to reach the Earth from distances as large as that to the GC.  
These neutrons could be produced in hardonic collisions of cosmic rays
(Medina-Tanco \& Watson~2001, Takahashi \& Nagataki~2001) which have been 
accelerated: (1) by a massive black hole associated with the Sgr A$^*$ 
(Levinson \& Boldt~2002); 
(2) by very young pulsars (Blasi, Epstein \& Olinto~2000, Giller \& Lipski~2002, 
Bednarek~2002); or 
(3) by shock waves of supernovae exploding into their own stellar winds
(Rhode, Enslin \& Biermann~1998). Relativistic protons, likely
responsible for injection of these neutrons, escape from the source and may 
reach the Earth after propagation in regular and  turbulent galactic magnetic fields.

In this paper we concentrate on the details of propagation of protons with energies
$\sim 10^{18}$~eV from a source located in the general direction of the GC (but at 
different distances from the Earth) applying two models of the galactic 
magnetic field in the Galactic Plane (GP) and halo. The existence of a large 
magnetic halo extending several kpc out of the GP in the perpendicular direction 
$z$ is suggested by the lack of cosmic ray deficit out of the GP
in the AGASA data at $\sim 10^{18}$ eV (see Clay~2001) and by observations of 
nearby spiral galaxies, e.g. NGC 253 - $z > 10$ kpc, $B\sim 7\mu$G 
(Beck et al.~1994),
NGC 4631 - $z > 8$ kpc, $B = 2\mu$G (Golla \& Hummel~1994), 
NGC 891 and NGC 4561 - $z\sim 3$ kpc, $B = 1\mu$G (Sukumar \& Allen~1991). 

The propagation effects of charged particles with extremely high 
energies (EHE) through the galactic regular magnetic field 
has been already analysed by many groups (e.g. Giller et al.~1994,
Stanev~1997, Zirakashvili et al.~1998; Biermann et al.~2000, Harari et al.~2000
Alvarez-Muniz et al.~2001, and references therein). 
In these papers the method of solving the "inverse" problem 
has been applied i.e. calculating trajectories of oppositely charged particles 
emitted from the observation point (Karaku\l a et al.~1972). 
This method is useful when considering many sources in the Galaxy but it is not 
better than the straightforward following the particle trajectory from its
source to the observer, when one point source is considered. The propagation of 
particles accelerated in the GC has been also studied recently by assuming that 
the field is axially symmetric allowing for the collection of all particles 
arriving in an annular stripe with radius equal to the distance of the Sun from the GC 
(e.g. Giller \& Zieli\'nska~2000, 
O'Neill, Olinto \& Blasi~2001). Such method, however, can not be applied 
to the propagation calculations from a source 
located at an arbitrary site in the Galaxy because the whole system 
is not symmetric then. Therefore the propagation of particles with energies 
$\sim 10^{18}$ eV from a discrete source situated in an arbitrary place in the 
Galaxy requires calculating trajectories of particles injected by the source and 
registering those intersecting a vicinity of the Earth. 
Such studies of CR propagation from the source located in or close to the 
GC in the context of the above mentioned observational results were carried out
by Clay~(2000), Clay et al.~(2000), Bednarek, Giller \& Zieli\'nska~(2001). 

The cosmic ray anisotropies obtained from the above studies, although strongly dependent
on the magnetic field adopted, have shown that it is very unlikely that a source of heavy 
nuclei with 
$E\cong 10^{18}$ eV would produce a compact excess on the sky, unless very close to us.
Therefore, we will consider here the possibility that the observed excess is due to cosmic 
ray protons and study their propagation from some point sources located in the direction 
of the GC.
In particular, we have studied the point source image as a function of time. 
We have also studied its dependence on the model of the regular magnetic field
in the Galaxy as well as on the proton energy. We have also considered whether the 
observed excess near the GC could be caused by particle emission from a single pulsar.
In a future paper we shall discuss the propagation from sources located in the Cygnus 
region (second excess found by the AGASA group) and a source at the 
Galactic anticentre to find out if a relatively close single pulsar of the Crab 
type can contribute significantly to the observed cosmic rays at EeV energies.  
Note that a dominant contribution of a single source to the cosmic ray spectrum
at lower energies $\sim 10^{15}-10^{16}$ eV (the knee region) has been 
suggested by Erlykin \& Wolfendale~(1997).

\section{Sources of EHE particles within the Galaxy}

We assume that particles with different energies are ejected isotropically 
by a short lived source, most likely a very young pulsar. Such very young 
pulsars (with milisecond periods) are presumably formed during the 
supernova type Ib/c explosions. The precursors of these supernova types are 
probably low mass Wolf-Rayet or oxygen-carbon stars rotating very fast and 
having light envelopes after explosion. Recent observations of diffuse hot plasma 
emitting X-rays (Yamauchi et al.~1990) suggest that in the past $10^5$ years 
about $10^3$ supernovae have exploded in the GC. Let's assume that at least one 
of them had parameters allowing acceleration of protons to energies above 
$10^{18}$ eV.

We follow the suggestion that pulsar winds are able to 
accelerate particles to energies $E$ corresponding to the full potential drop 
available across the polar cap region (Gunn \& Ostriker 1969, Blasi, 
Epstein \& Olinto 2000), 
\begin{eqnarray}
E = {{eBR^3\Omega^2}\over{c^2}} \approx 
{{6.6\times 10^{19}B_{13}}\over{P_{\rm ms}^2}}~~{\rm eV},
\label{row1}
\end{eqnarray}
where $\Omega = 2\pi/P$, $P = 10^{-3}P_{\rm ms}$ s is the pulsar period, 
$R = 10^6$ cm is the radius of the neutron star, $B = 10^{13}B_{13}$ G 
is its surface magnetic field, $e$ is the elementary charge, and 
$c$ is the velocity of light. Eq.~(\ref{row1}) allows us to constrain the 
parameters of the pulsar able to accelerate protons to energies 
$E \ge 10^{18}$ eV. The following condition has to be fulfilled
\begin{eqnarray}
P_{\rm ms}\le 8B_{13}^{1/2}. 
\label{row2}
\end{eqnarray}
\noindent
If the pulsar loses its rotational energy, $E_{\rm rot}$, only on 
emission of the dipole radiation with the power $L$, then its
period at specific time $t$ is determined by the equation 
\begin{eqnarray}
{\dot E}_{\rm rot} = L~~~~~~~~ {\rm \Longrightarrow} ~~~~~ 
I\Omega\dot{\Omega} = -B^2R^6\Omega^4/6c^3
\label{row3} 
\end{eqnarray}
\noindent
where $I = 1.4\times 10^{45}$ g cm$^{-2}$ is the neutron star
moment of inertia. As a result of this energy losses the period of the pulsar 
changes in time according to
\begin{eqnarray}
P_{\rm ms}^2(t) = P_{0,ms}^2 + 3.5\times 10^{-8} B_{13}^2t,
\label{row4} 
\end{eqnarray}
\noindent
where $t$ is in seconds, and $P_{0,ms}$ is the pulsar initial period.
By reversing Eq.~(\ref{row4}) and applying Eq.~(\ref{row2}), we estimate 
the time elapsed from the pulsar formation, $t_{\rm acc}$, during 
which protons will be accelerated above $10^{18}$ eV
(assuming that $P_{0,ms}\ll P_{\rm ms}(t_{\rm acc})$). It is 
\begin{eqnarray}
t_{\rm acc}\approx 60 B_{13}^{-1}~~{\rm yr}.
\label{row5} 
\end{eqnarray}
\noindent
Since this time is relatively short, when compared to the time scale of 
particle propagation, we can consider such 
injection of relativistic protons as instantaneous. 

The hypothesis for the origin of particles with such energies in pulsars 
makes the model flexible since, in principle, the pulsar can be born at an 
arbitrary site in the Galaxy. Such proposition can give a better explanation of 
the observations (i.e. the second AGASA excess in the Cygnus region) and allows
investigation of particle propagation from the source 
shifted from the GC itself (see Bellido et al. 2001), or 
in the direction of the GC but located at a different distance. 
Since the propagation of protons with EeV energies
in the Galactic magnetic field takes tens to hundreds of thousand years
longer than that of radiation, this pulsar may not
necessarily be visible presently as a source of $\gamma$-rays, neutrinos or
neutrons, unless it is immersed in a dense molecular cloud accumulating protons
(a model recently considered by Bednarek~2002). Almost instantaneous injection of 
$\sim 10^{18}$ eV protons by such pulsar together with a particular structure of 
the regular magnetic field in the Galaxy may result in quite unusual images of 
the source at different times after injection. 

\section{The structure of the Galactic magnetic field}

Protons ejected from a point-like source propagate in the Galactic magnetic 
field which consists of regular $B_{\rm reg}$ and irregular $B_{\rm irr}$ 
components. We have adopted two different models for the regular Galactic 
magnetic field. The first model (our model I)
proposed and described in detail by Urbanik, Elstner \& Beck~(1997) 
bases on observations of near-by spiral galaxies. 
$B_{\rm reg}$ has a toroidal component, confined
mainly to the disk, and a large scale poloidal component, extending up to 
$z = 10$ kpc. The total $B_{\rm reg}$ does not exceed 2$\mu$G anywhere. For the 
detailed structure
of the magnetic field in the halo see Fig.~2 in the Urbanik et al. paper.

As the second possibility (our model II) we adopt the bisymmetric field model 
with field reversals 
and odd parity (BSS-A) proposed by Han \& Qiao~(1994) and applied for the
particle propagation purposes by e.g. Stanev ~(1997). 
This model incorporates the knowledge from experimental observations of our Galaxy 
as well as many other galaxies. It consists of the GP component  
in which the field strength at a point (r, $\theta$) is described by 
\begin{eqnarray}
B(r, \theta) = B_0(r) \cos[\theta - \beta \ln (r/r_0)], 
\label{row6}
\end{eqnarray}
\noindent
where $r_0$ is the 
Galactocentric distance of the location with maximum field strength at  $l = 0^{\rm o}$,
$\beta = -5.67$, and  $r_0 = 10.55$ kpc. $B_0(r)$ is taken to be 
$3R_{\rm GC}/r$ $\mu$G above $r = 4$ kpc and constant below this distance from 
the centre of the Galaxy, where $R_{\rm GC} = 8.5$ kpc. The radial and azimuthal 
components of the magnetic field in the halo are described by
\begin{eqnarray}
|B(r, \theta, z)| = |B(r, \theta)| exp (-|z|/z_0),
\label{row7}
\end{eqnarray}
\noindent
with two scale heights $z_0 = 1$ kpc for $|z| < 0.5$ kpc and 
$z_0 = 4$ kpc for $|z| > 0.5$ kpc with the field direction at the disk crossing 
unchanged (see Stanev 1997). The magnetic field component perpendicular to the 
GP, $B_{\rm z}$, is assumed constant with the value of 0.3$\mu$G and 
is always directed to the north.
 
In both models we describe the irregular field, $B_{\rm irr}$, as a sum of many 
plane
waves with isotropically distributed wave vectors and amplitudes corresponding to
the Kolmogorov power spectrum. Its mean value is 2 $\mu$G in the disk with
$z\pm 500$ pc, and 0.5 $\mu$G in the spherical halo with radius 20 kpc.
The irregularity scale is, however, different in the disk and the halo: the
longest wave is 150 pc in the disk and 7 kpc in the halo.

\section{Propagation of EHE protons}

We calculate numerically the proton trajectories within the range of 
energies corresponding to those of the AGASA excess
($10^{17.9}-10^{18.3}$~eV) and the SUGAR excess ($10^{17.9}-10^{18.5}$~eV)
from the direction of the GC. For $1.2\times 10^6$ protons ejected 
isotropically from a point-like source located at three different
points in the Galaxy we record the parameters (numbers, directions) of particles
intersecting a sphere with the radius of 250~pc centred on the Earth.  
These events are considered as observed by a detector on the Earth.
Other nuclei with energy $E$ and charge $Z$, propagating in the 
magnetic field by a factor $\alpha$ stronger behave 
exactly as protons with energies $E/(\alpha Z)$.  

We consider the sources of charged particles located: (a) exactly at the GC 
at the distance of 8.5 kpc from the Sun; 
(b) 2 kpc from the Sun towards the GC (for this distance the potential 
source of the SUGAR excess lays still within the region of the galactic disk);
(c) 8.5 kpc from the Sun towards the direction of the SUGAR excess 
displaced from the GC direction by
$\sim 7.5^{\rm o}$, and $\sim 2.5^{\rm o}$ below the GP 
(about $\sim 400$ pc from GP). Below, we discuss the 
calculation results for these three cases.

\subsection{A source in the Galactic Centre}

The GC seems to be one of the most likely 
site for particle acceleration to energies above $\sim 10^{18}$~eV. In Sect.~2 
we argue that the best candidate source is a very young neutron star with a 
milisecond period and surface magnetic field $B\cong 10^{13}$ G. 
Since the particles can be acclerated by such a source for 
a short time, we assume the instantaneous injection of protons with 
energies $1,2,3\times 10^{18}$ eV.
 
First we discuss the results of calculations of proton trajectories from a 
point-like source exactly in the GC for the two galactic field models.
The numbers of particles arriving at different times after injection (the 
time of flight along the straight line being subtracted) to the sphere 
centered on the Earth are displayed in the form of histograms in Figs.~1h, 2h, and 3h 
for model I, and in Figs.~1p, 2p, and 3p
for model II. It becomes evident that the distribution of the arrival 
times of particles with energies by a factor 2-3 larger is completely different.
The bulk of particles arrive to the observer within $\sim 2.5\times 
10^4$ years (model I) and $5\times 10^4$ years (model II) for $3\times 10^{18}$ eV 
(Figs.~1h and 1p), up to $\sim 10^5$ years (model I) and 
$10^6$ years (model II) for $10^{18}$ eV (Figs. 3h and 3p). 

In Figs.~1,2, and 3 from a) to f) (model I) and from i) to n) (model II) we 
show maps (in galactic coordinates, with longitude increasing to the left) with 
the arrival directions of protons intercepting the sphere around the Earth within 
consecutive time delay intervals chosen accordingly to the particle energy and 
magnetic field model (see figure captions). Maps summed up over time are in 
Figs. g) and o) showing the direction distribution in the case of 
a steady source. 

Let's first concentrate on the results for model I. The most interesting 
feature for protons with energies $(2 - 3)\times 10^{18}$ eV 
is the particle clustering in multiple images of the source.
These images appear at different places on the sky at different times 
after injection. A large number of protons reach the 
Earth's vicinity from directions close to the GC (shifted by about 
$\sim 10^{\rm o}$ towards positive longitudes) creating an extended source
with the radius of about $20^{\rm o}-30^{\rm o}$. {\it Note that this location is 
consistent with the direction towards the excess of particles obtained 
by AGASA and SUGAR. Therefore,  
for the galactic magnetic field structure proposed by Urbanik et 
al. the source of these particles can be actually located exactly in the 
Galactic Centre !}

For protons with energies $10^{18}$ eV the arrival directions become 
much more scattered
(see Figs.~3a to 3h). In this case, protons arrive from a large part 
of the sky, almost independently of time after injection, apart from the 
peak for the first $\sim 2\times 10^4$ years. Increasing slightly 
the field strength will cause the discovered features shifting to higher  
energies and fitting better to the energy range where the actual 
excess of particles has been detected.

The arrival directions of protons and their time distributions are completely 
different for the field model II. There are no protons arriving directly 
from the actual position of the source at the GC (out of $\sim 10^6$ ejected)
up to $3\times 10^{18}$~eV. For protons with $3\times 10^{18}$~eV, 
only a single image of the source
is visible at a high negative latitude. It is created mainly by protons arriving 
with relatively small time delay with respect to the rectilinear propagation, 
i.e. 
within less than $2\times 10^4$ years (see Fig.~1p). For lower proton energies the 
image of the source is also centred on high galactic latitudes becoming 
broader and stronger. Particles arrive to the Earth much later than for model I i.e.
after $(2 - 4)\times 10^5$ years for $2\times 10^{18}$~eV and $(2 - 7)\times 10^5$ years for
$3\times 10^{18}$~eV. In spite of the instantaneous injection the anisotropy due to the 
source would be visible in the same directions for a long time.

By comparing our calculation results on the proton anisotropy for the two 
magnetic field models we conclude that the propagation of charged particles 
is very sensitive to their energies and to the structure and strength of the 
magnetic field. The two models, both based on experimental observations, give 
totally different predictions concerning the particle angular distribution on 
the sky, meaning that one should be very careful with drawing any conclusions 
based on one particular model of the Galactic magnetic field. 

\subsection{A source at 2 kpc towards the Galactic Centre}

The distance to the source of particles responsible for the AGASA-SUGAR
excess can not be constrained by the observations. Therefore it is
reasonable to investigate the case of a source located much closer than the GC. 

In Figs.~4,5 and 6 we show maps of the arrival directions of protons 
intercepting the sphere around the Earth for a source at 2 kpc. 
The maps summed up over time are shown in Figs. g) and o). 
For this relatively small distance in all considered cases (different proton 
energies 1,2, and $3\times 10^{18}$ eV and both structures of the galactic 
magnetic field) a large number of protons reach the Earth 
by moving almost along straight lines (see initial peaks in histograms 
showing the time distribution of the arriving particles). Therefore the excess 
of relativistic protons should be observed for a short time only. 

In model I protons injected with $3\times 10^{18}$ eV can reach 
the Earth only sporadically, if the time delay after injection is large 
(see Figs.~4 b-f and h). However, the 
number of the delayed protons quickly increases with decreasing energy and at 
$10^{18}$ eV the particles are arriving to us for 
$\sim 2\times 10^5$ years. The intensity of these 
particles differ significantly at specific time intervals showing several, periodic, 
maxima, seen clearly in Figs.~5h and 6h. The magnetic field $B_{\rm reg}$ in the halo 
turns out to be directed from the Sun towards the source, so that this periodic 
arrival corresponds to the particle multiple gyro-orbits.
During the maximum intensities particles arrive from a great circle on the sky, roughly 
perpendicular to the GP and crossing it at $l\sim 90^{\rm o}$ and $270^{\rm o}$.

The maps of arrival directions and their time structure differ significantly 
for model II also for this case when the source is 
relatively close to us (2 kpc). There is clear 
deflection of the instantaneuos image of the source towards the south,  
increasing for lower energies.
For $3\times 10^{18}$ eV, there exists a secondary 
extended image shifted by a large angular
distance towards negative galactic 
longitudes at the time interval $(1.3 - 2)\times 10^5$ years (see Figs~4k,l, 
and p). For lower energies (2$\times 10^{18}$ eV) a large fraction of
protons are delayed by up to $\sim 10^6$ years after injection. 
They arrive mainly from the southern hemisphere due to the asymmetric form of 
the $B_{\rm z}$ component in the halo. An interesting thing is that there is an 
elongated image in the directions opposite to that of the source ($l\sim 180^{\rm o}$,
$\delta < 0$).

\subsection{A source towards the AGASA-SUGAR excess}

As we mentioned above the excess seen in the SUGAR data is consistent with a
point-like source but displaced from the GC direction by
$\sim 7.5^{\rm o}$. The maximum probability map of the AGASA excess is shifted 
from the GC direction even more. 
Therefore, it is reasonable to consider a source located at a certain distance
from the exact position of the GC. We have calculated trajectories of 
isotropically injected protons assuming that the source is 
8.5 kpc away but in the direction of the SUGAR excess. As before we 
collect protons with energies $1,2,4,10\times 10^{18}$ eV intersecting the sphere
 around the Earth for the two magnetic 
field models. Now the source is located about $\sim 400$ pc below the GP. 
The obtained maps (integrated over time) are presented in Fig.~7.
However,  there are no significant differences in comparison
to the calculations from the source located exactly in the GC. 
Nevertheless, the image sizes for both cases are rather large at the considered 
energies and would be close to the AGASA excess size for $E \geq 4\times 
10^{18}$ eV.

\section{Single pulsar as a plausible source}

Let us consider whether a single pulsar could be responsible for the observable excess
from the GC. The power emitted by the pulsar accelerating protons can be expressed 
by the proton energy $E$ from Eqs.~\ref{row1} 
and~\ref{row3}, 
\begin{eqnarray}
L = c E^2/6e^2\approx 
5.5\times 10^{40} E_{18}^2~{\rm erg~s}^{-1},
\label{row8}
\end{eqnarray}
\noindent
where $E = 10^{18}E_{18}$ eV. It is difficult to state 
at present if such powerful pulsar is present at the GC. One remnant
of a young supernova with age of $\sim 80$ years (G0.570-0.018) has been
recently reported by Senda et al.~(2001). Also a strong $\gamma$-ray source 
($\sim 2\times 10^{37}$ erg s$^{-1}$) with a very hard spectrum is observed
by the EGRET instrument from the GC (Mayer-Hasselwander et al.~1998).
The nature of this source is at present unknown.

The rate, $r_{\rm obs}$, at which protons arrive to the Earth 
(i.e. intersect the sphere with 
the radius $R_{\rm E} = 250$ pc) from the 'AGASA-SUGAR source' can 
be estimated from the flux $F_{\rm obs} = 9\times 10^{-14}$ m$^{-2}$ s$^{-1}$ 
derived from the SUGAR observations. It is 
\begin{eqnarray}
r_{\rm obs} =  \pi R_{\rm E}^2 F_{\rm obs}\approx 1.7\times 10^{25} 
{\rm particles}~~s^{-1}. 
\label{row9}
\end{eqnarray}
On the other side
the number of particles $\Delta N$ injected by the pulsar with energies between
$E_1$ and $E_2$ would be 
\begin{eqnarray}
\Delta N = {{c^2 \eta I}\over{Ze R_{\rm NS}^3 B}} \ln(E_2/E_1) \approx 
{{2.6\times 10^{44}\eta}\over{B_{13}}}\ln{{E_2}\over{E_1}}~~{\rm protons},
\label{row10}
\end{eqnarray}
\noindent
where $Ze$ is the particle charge, and $\eta$ is the efficiency of 
particle acceleration defined as the ratio of the number of injected particles 
to that of the maximum number possible given by the 
Goldreich \& Julian~(1969) density at the pulsar light cylinder.
Our present calculations have shown that the source image changes significantly even if
proton energy goes from 3 to $4\times 10^{18}$ ev, i.e. changes by 30\%. Therefore,
we think that it is quite likely that the observed excess is actually due to a narrower
energy band than the quoted factor of 4 (unless the source is quite close to us) and we 
adopt that $E_2/E_1 = 1.2$ (with $E\cong 3\times 10^{18}$ eV). 
Then we have
\begin{eqnarray}
\Delta N \approx 
{{5\times 10^{43}\eta }\over{B_{13}}}~~{\rm protons},
\label{row11}
\end{eqnarray}
\noindent
For a pulsar located in the GC the fraction $f$ of these particles giving the image 
close to the source (for model I, Fig 1h) is $\sim 5\times 10^{-4}$ 
(about 550 particles out of $1.2\times 10^6$) i.e. 2.5 times larger that that for  
a straightforward propagation. As these particles arrive to the Earth vicinity within 
$\Delta t\cong 5\times 10^3$ years, we should expect for their average rate 
\begin{eqnarray}
r_{\rm 1} =  f {{\Delta N}\over{\Delta t}} \sim 2\times 10^{29}
{{\eta}\over{B_{13}}} {\rm protons}~~s^{-1}. 
\label{row12}
\end{eqnarray}
\noindent
Comparing this with $r_{\rm obs}$, we obtain that $\eta/B_{13}\sim 10^{-4}$ for this case
(a value probably not unreasonable !).  

It may be, however, that the observed excess is due to a more delayed image of a source 
located somewhere else. Let's assume (somewhat arbitrarily) that we can apply our 
calculation results to this case as well. For example, the secondary compact images 
(Fig. 1b,c,d,e) are visible for $\Delta t\cong 2\times 10^4$ years and the fraction
of the emitted particles producing them equals $f\approx 7\times 10^{-5}$ 
($\sim 80$ particles per one secondary image, out of $1.2\times 10^6$). 
Then their average 
rate is
\begin{eqnarray}
r_{\rm 2} \cong 5\times 10^{27}
{{\eta}\over{B_{13}}} {\rm protons}~~s^{-1}, 
\label{row13}
\end{eqnarray}
\noindent
and, to agree with $r_{\rm obs}$,~~$\eta/B_{13}\cong 3\times 10^{-3}$.
If the pulsar was closer then the efficincy $\eta$ would have to be smaller (Figs. 4,5,6).

We can not, however, exclude a possibility that the observed excess is being produced
by even more delayed protons with a smaller intensity and a longer arrival time 
$\Delta t$.
If the total observed CR flux was not produced by pulsars (but by some other sources)
then such a bunch of particles could produce an increase on the average CR flux.

For the bisymmetric field model proposed by Han \& Qiao~(1994) the image of the source
located at the GC does not appear in the direction to the real source up to 
$3\times 10^{18}$ eV, but is shifted from its real position by a 
large angle. Therefore, in this case protons can not be responsible for the 
observed excess.  

\section{Conclusions}

The main conclusions of our calculations of the propagation of protons with 
energies $\sim 10^{18}$ eV, injected by an isotropic point-like source 
in the direction towards the Galactic Centre are the following:

\begin{itemize}

\item  Protons injected instantaneously by a point-like source at the 
Galactic Centre, arriving to the Earth after propagation in the galactic 
regular and irregular magnetic fields, can form multiple images at directions
completely different from those towards the source, as well as images shifted only slightly 
from the position towards the source, and large scale anisotropies (north-south 
asymmetry) depending on the proton energies and the time elapsed from injection.
These multiple images appear only at relatively {\it narrow energy 
ranges} of particles between $(1 - 3)\times 10^{18}$ eV (for our magnetic field
models). The arrival directions 
of protons with lower energies are broadly scattered.
Particles with higher energies create an extended image centered on the source. 

\item The results of particle propagation for the two considered 
magnetic field models
give totally different predictions for the particle angular distribution 
on the sky, meaning that one should be very careful when drawing conclusions 
based on one particular model.

\item The application of the Han \& Qiao magnetic model  
produces strong north-south anisotropy of particles with energies 
$\sim (2 - 3)\times 10^{18}$ eV. Most of the delayed particles arrive 
from the southern Galactic hemisphere, in contrast to the models of 
extragalactic origin of the highest energy cosmic rays favouring the 
northern hemisphere (e.g. Biermann et al.~2000). 

\item Our results do not depend on the particular distribution of the irregular 
magnetic component, providing that its magnitude is not larger than assumed here. 
A detailed experimental study of the $\sim 10^{18}$ eV CR anisotropy
should give information about the validity of the pulsar model and/or that of the 
large scale galactic magnetic field.

\item It is interesting to note that 
the maximum proton fluxes predicted for a pulsar model are larger than that observed.
Thus, the model of isotropic and instantaneous particle injection by a 
pulsar, located close to the GC, could explain the observed flux of particles in 
the AGASA-SUGAR excess as due to prompt protons (travelling almost along 
straight lines) or as due to the delayed protons. In the last case, however, the 
source would be in some other direction.

\end{itemize}

\section*{Acknowledgments.}
\noindent
This work is supported by the KBN grants No. 2P03C 006 18 and 5P03D 025 21.

\newpage

\newpage

%
\begin{figure*} 
  \vspace{15.5cm} 
 \includegraphics{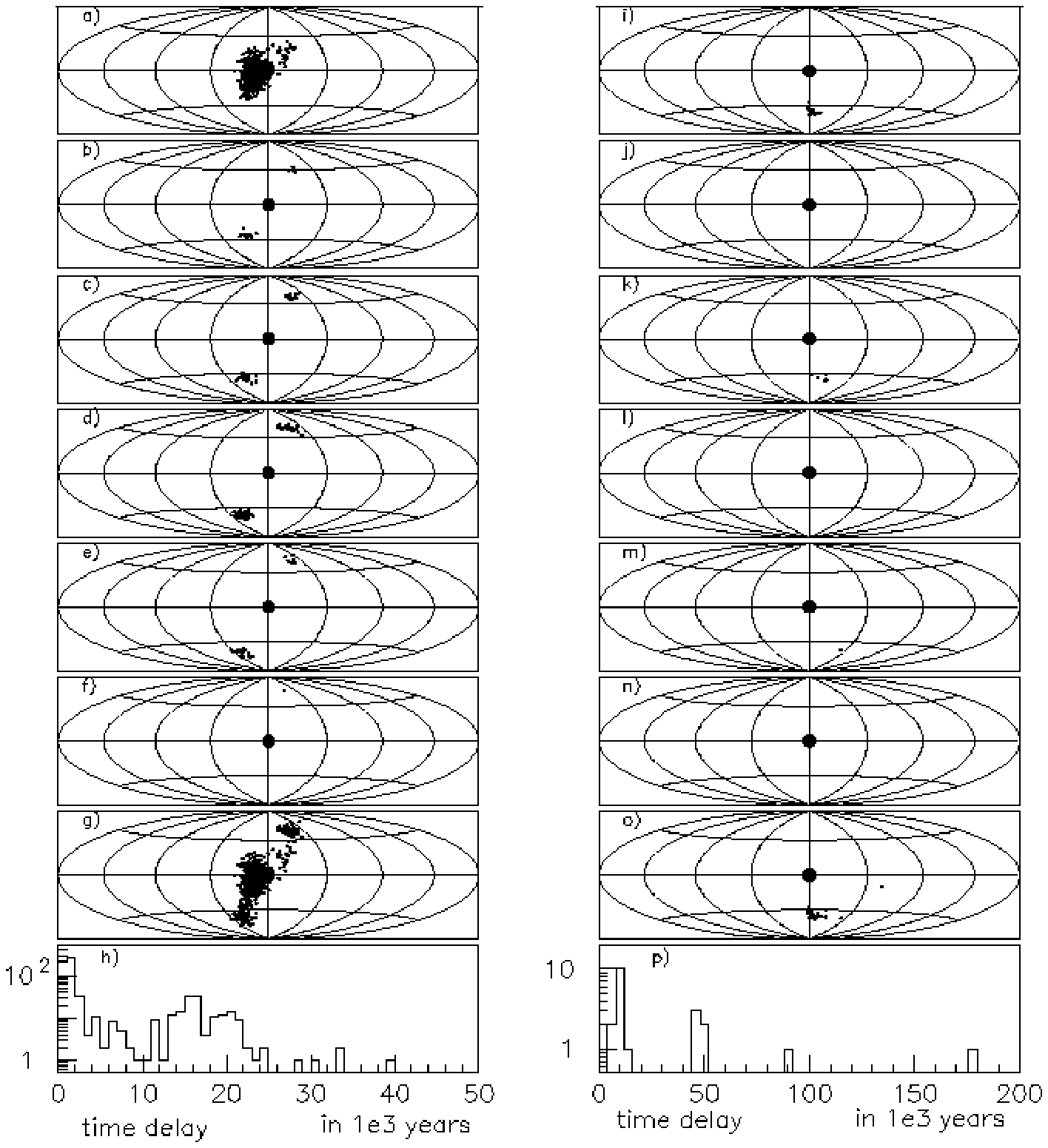}
 \caption[]{Arrival directions of protons with energies $3\times 10^{18}$~eV
injected by a point source in
the GC (marked by the large dot) for model I
(left) and model II (right). Maps (in galactic coordinates)
from a) to f) show directions of particles arriving in consecutive time
delay intervals of $5\times 10^3$~yr, i.e.  a) is for $0 - 5\times 10^3$~yr,
 ... , f) $(2.5 - 3)\times 10^4$~yr, and
from i) to o) with intervals of $2\times 10^4$~yr, i.e.  i) is for $0 - 2\times 
10^4$~yr, ... , o) $(1 - 1.2)\times 10^5$~yr. 

g) and o) Arrival directions integrated over time. 

h) and p) Delay time distribution of arriving particles; time in units of $10^3$~yr.}
   \label{fig1}  
\end{figure*} 
%
%
\begin{figure*} 
  \vspace{15.5cm} 
 \includegraphics{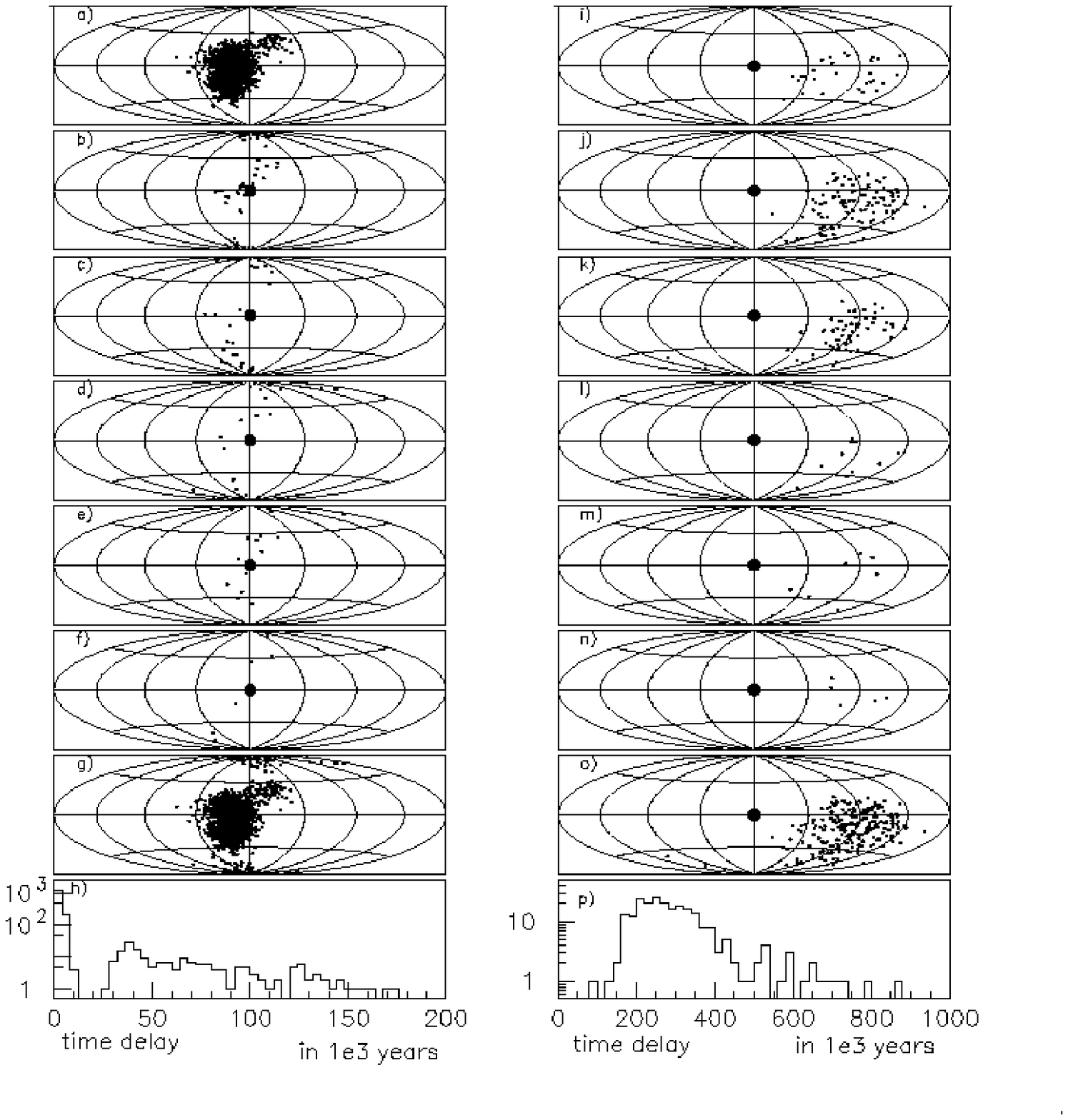}
 \caption[]{As in Fig.~1 but for protons with energies $2\times 10^{18}$
eV. Maps for model I are for time delay intervals $2\times 10^4$ yr, 
for model II - $10^5$ yr. Maps for model II start from $10^5$ yr.}
    \label{fig2}  
\end{figure*} 
%

%
\begin{figure*} 
  \vspace{15.5cm} 
 \includegraphics{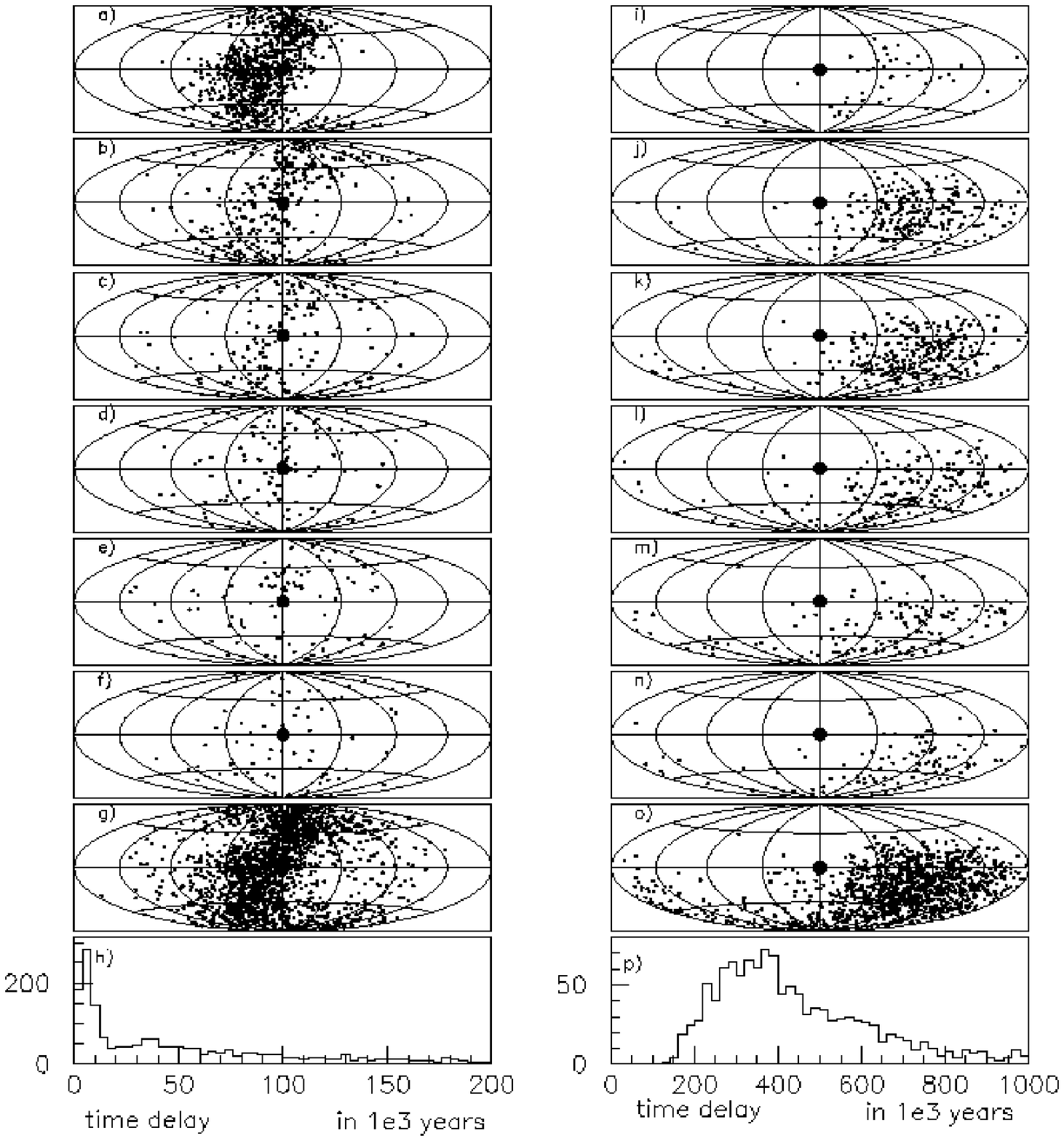}
 \caption[]{As in Fig.~1 but for protons with energies $10^{18}$ eV.
Maps for model I are for time delay intervals $2\times 10^4$ yr, 
for model II - $10^5$ yr. Maps for model II start from $10^5$ yr.}
    \label{fig3}  
\end{figure*} 
%

%
\begin{figure*} 
  \vspace{15.5cm} 
 \includegraphics{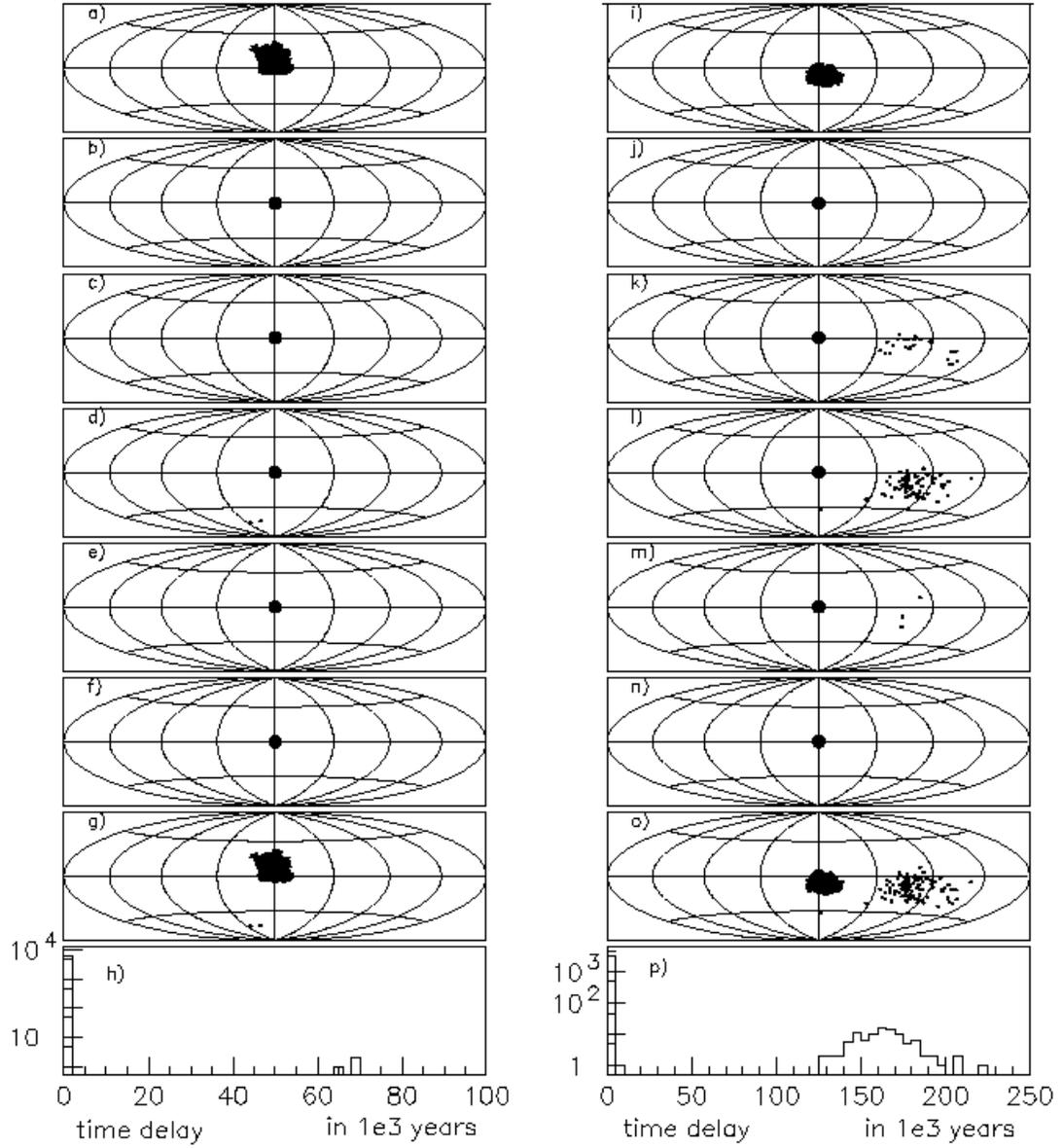}
 \caption[]{Arrival directions of protons with energies $3\times 10^{18}$ eV
injected by a point source at a distance of 2 kpc towards the direction of
the GC (marked by the large dot) for model I
(left) and model II (right). Time delay intervals are $2\times 10^4$ yr - for model I 
and $5\times 10^4$ yr - for model II.}
    \label{fig4}  
\end{figure*} 
%

%
\begin{figure*} 
  \vspace{15.5cm} 
 \includegraphics{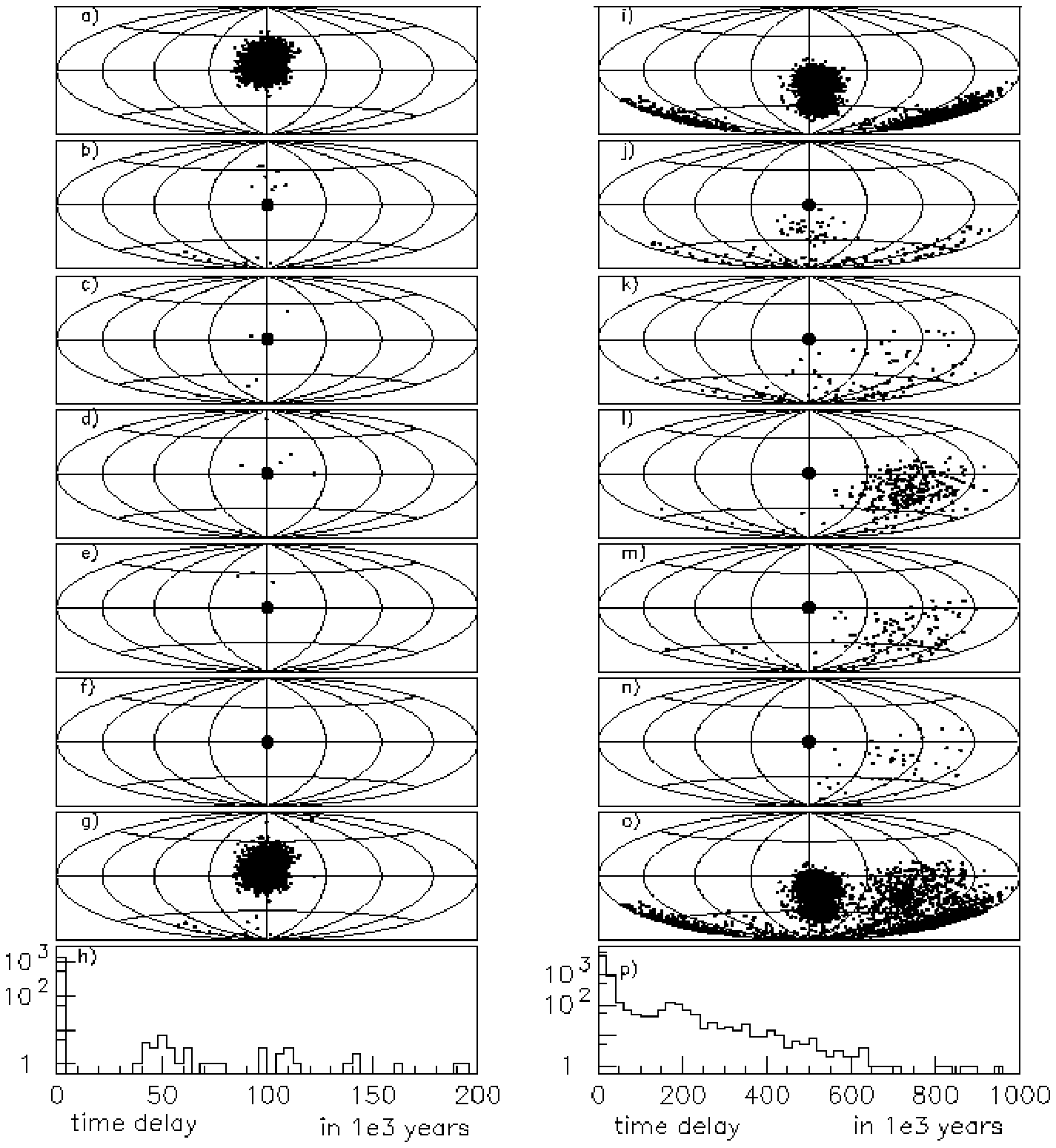}
 \caption[]{As in Fig.~4 but for protons with energies $2\times 10^{18}$
eV. Time delay intervals are $3\times 10^4$ yr - for model I 
and $5\times 10^4$ yr - for model II.}
    \label{fig5}  
\end{figure*} 
%

%
\begin{figure*} 
  \vspace{15.5cm} 
 \includegraphics{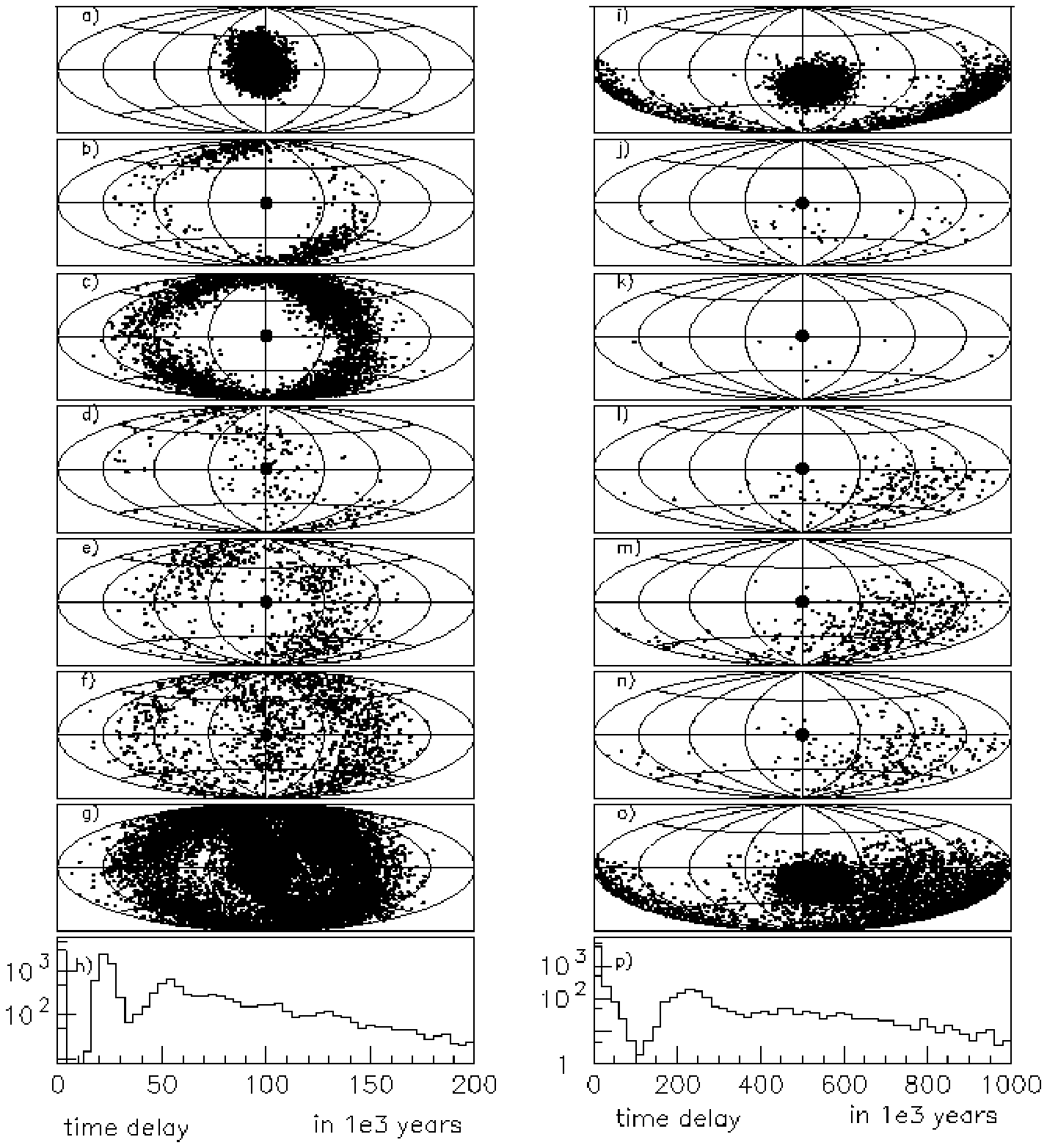}
 \caption[]{As in Fig.~4 but for protons with energies $10^{18}$ eV.
Time delay intervals are $10^4$ yr - for model I and $5\times 10^4$ yr - for 
model II.}
  \label{fig6}  
\end{figure*} 
%

%
\begin{figure*} 
  \vspace{9.cm} 
 \includegraphics{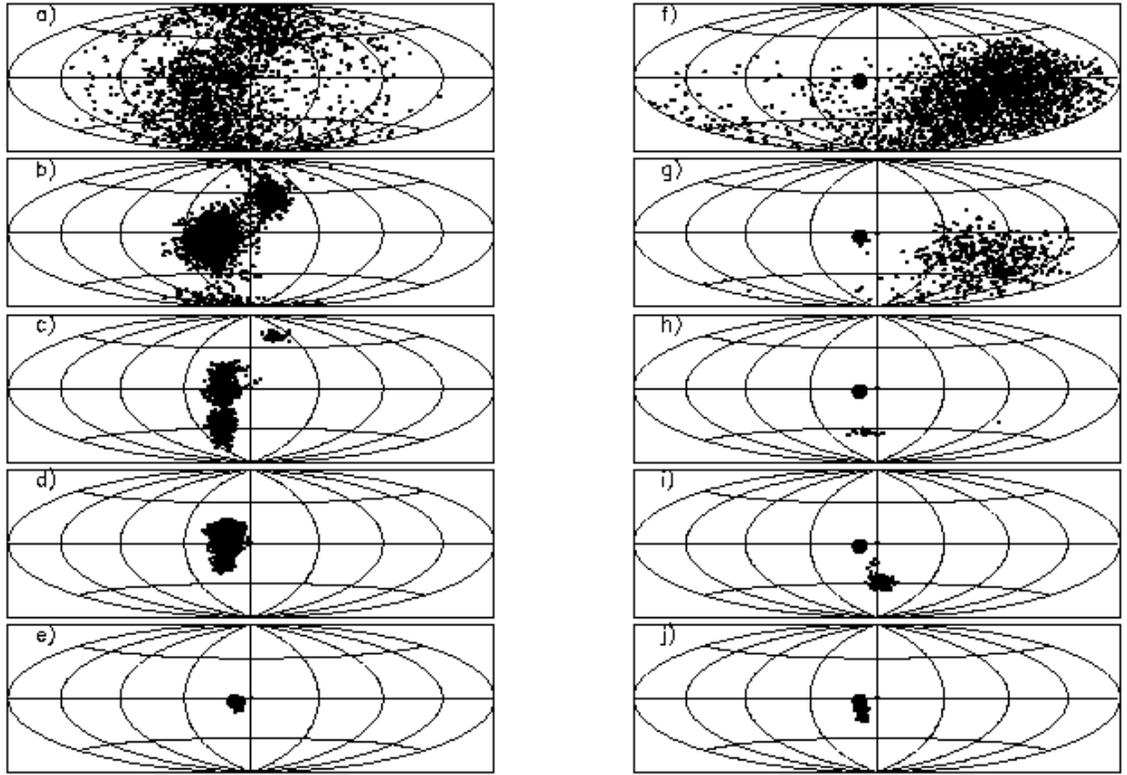}
 \caption[]{Arrival directions for all times of protons with energies 
$10^{18}, 2\times 10^{18}, 3\times 10^{18}, 4\times 10^{18}$ eV and $10^{19}$ eV 
(figures from top to bottom)
injected by a point source at a distance of 8.5 kpc towards the direction of
the SUGAR excess (marked by the large dot) for model I (left) 
and model II (right).}
  \label{fig7}  
\end{figure*} 
%

\end{document}